\documentclass[preprint,12pt]{elsarticle}
\usepackage{epsfig}
\usepackage{amssymb}
\usepackage{amsthm}
\usepackage{amsmath}
\biboptions{sort&compress}

\journal{Theoretical Computer Science}

\newtheorem{thm}{Theorem}
\newtheorem{lem}[thm]{Lemma}
\newproof{pf}{Proof}

\newcommand{\CC}{\mathcal{C}}

\begin{document}

\begin{frontmatter}

\title{Efficient sub-$5$ approximations for\\minimum dominating sets in unit disk graphs\tnoteref{ALL}}
\tnotetext[ALL]{An extended abstract of this paper appeared in the proceedings of the 10th Workshop on Approximation and Online Algorithms (WAOA'12), \emph{Lecture Notes in Computer Science} \textbf{7846} (2013), 82--92. Research partially supported by FAPERJ and CNPq grants.}

\author[gf]{G.D. da Fonseca}
\ead{fonseca@uniriotec.br}

\author[ch]{C.M.H. de Figueiredo}
\ead{celina@cos.ufrj.br}

\author[vs]{V.G. Pereira de S\'a}
\ead{vigusmao@dcc.ufrj.br}

\author[rm]{R.C.S. Machado}
\ead{rcmachado@inmetro.gov.br}

\address[gf]{Universidade Federal do Estado do Rio de Janeiro, Brazil}

\address[ch,vs]{Universidade Federal do Rio de Janeiro, Brazil}

\address[rm]{Instituto Nacional de Metrologia, Qualidade e Tecnologia, Brazil}

\begin{abstract}
A unit disk graph is the intersection graph of $n$ congruent disks in the plane. Dominating sets in unit disk graphs are widely studied due to their applicability in wireless ad-hoc networks. Because the minimum dominating set problem for unit disk graphs is \textbf{NP}-hard, numerous approximation algorithms have been proposed in the literature, including some PTASs.
However, since the proposal of a linear-time $5$-approximation algorithm in 1995, the lack of efficient algorithms attaining better approximation factors has aroused attention.
We introduce an $O(n+m)$ algorithm that takes the usual adjacency representation of the graph as input and outputs a $44/9$-approximation. This approximation factor is also attained by a second algorithm, which takes the geometric representation of the graph as input and runs in $O(n \log n)$ time regardless of the number of edges.
Additionally, we propose a $43/9$-approximation which can be obtained in $O(n^2 m)$ time given only the graph's adjacency representation. It is noteworthy that the dominating sets obtained by our algorithms are also independent sets.
\end{abstract}

\begin{keyword}
approximation algorithms; dominating set; unit disk graph.
\end{keyword}

\end{frontmatter}

\section{Introduction}

A \emph{unit disk graph} $G$ is a graph whose $n$ vertices can be mapped to points in the plane and whose $m$ edges are defined by pairs of points within Euclidean distance at most $1$ from one another. Alternatively, one can regard the vertices of $G$ as mapped to coplanar disks of unit diameter, so that two vertices are adjacent whenever the corresponding disks intersect.

A \emph{dominating set} $D$ is a subset of the vertices of a graph such that every vertex not in $D$ is adjacent to some vertex in $D$. An \emph{independent set} is a subset of mutually non-adjacent vertices. An \emph{independent dominating set} is a dominating set which is also an independent set. Note that any maximal independent set is an independent dominating set.

Dominating sets in unit disk graphs are widely studied due to their application in wireless ad-hoc networks~\cite{heuristics}. Since it is \textbf{NP}-hard to compute a minimum dominating set of a unit disk graph~\cite{udg}, several approximation algorithms have been proposed~\cite{cccg,erlebach10,esa-Gibson,ptas-geometric,heuristics,ptas-graph-journal,zou11}. Such algorithms are of two main types. \emph{Graph-based algorithms} receive as input the adjacency representation of the graph and assume no knowledge of the point coordinates, whereas \emph{geometric algorithms} work in the Real RAM model of computation and receive solely the vertex coordinates as input\footnote{The Real RAM model is a technical necessity, otherwise storing the coordinates of the vertices would require an exponential number of bits~\cite{integer}.}. 

If the coordinates of the $n$ disk centers are known, the $m$ edges of the corresponding graph $G$ can be figured out easily. It can be done in $O(n+m)$ time under the Real RAM model with floor function and constant-time hashing, and in $O(n \log n + m)$ time without those operations~\cite{bentley}. Thus, for the price of a conversion step, graph-based algorithms can be used when the input is a unit disk realization of $G$. However, unless \textbf{P}=\textbf{NP}, no efficient algorithm exists to decide whether a given graph admits a unit disk realization~\cite{breu}, let alone exhibit one. As a consequence, geometric algorithms cannot be efficiently transformed into graph-based algorithms. In this paper, we introduce approximation algorithms of both types, benefiting from the same approximation factor analysis. The proposed graph-based algorithm runs in $O(n+m)$ time, and the geometric algorithm runs in $O(n \log n)$ time regardless of $m$.

\paragraph*{Previous algorithms}
A graph-based $5$-approximation algorithm that runs in $O(n+m)$ time was presented in~\cite{heuristics}. The algorithm computes a maximal independent set, which turns out to be a $5$-approximation because unit disk graphs contain no $K_{1,6}$ as induced subgraphs, as shown in that same paper.\footnote{The graph $K_{1,q}$ consists of a vertex with $q$ pendant neighbors.}

Polynomial-time approximation schemes (PTAS) were first presented as geometric algorithms~\cite{ptas-geometric} and later as graph-based algorithms~\cite{ptas-graph-journal}. Also, a graph-based PTAS for the more general disk graphs was proposed in~\cite{esa-Gibson}. Unfortunately, the complexities of the existing PTASs are high-degree polynomials. For example, the PTAS presented in~\cite{ptas-graph-journal} takes $O(n^{225})$ time to obtain a $5$-approximation (\mbox{using} the analysis from~\cite{cccg}). Although its analysis is not tight, the running time is too high even for moderately large graphs. The reason is that these PTASs invoke a subroutine that verifies by brute force whether a graph admits a dominating set with $k$ vertices. The verification takes $n^{O(k)}$ time, and it is unlikely that this can be improved (unless \textbf{FPT}=\textbf{W[1]}, as proved in~\cite{marx}). Such a subroutine is applied to several subgraphs, and the value of $k$ grows as the approximation error decreases. A similar strategy was used in~\cite{ids-ptas} to obtain a PTAS for the minimum independent dominating set.

The lack of fast algorithms with approximation factor less than $5$ was recently noticed in~\cite{cccg}, where geometric algorithms with approximation factors of $4$ and $3$ and running times respectively $O(n^{9})$ and $O(n^{18})$ were presented. While a significant step towards approximating large instances, those algorithms require the geometric representation of the graph, and their running times are still polynomials of rather high degrees. Linear and near-linear-time approximation algorithms constitute an active topic of research, even for problems that can be solved exactly in polynomial time, such as maximum flow and maximum matching~\cite{maxflow,matchings}.

It is useful to contrast the minimum dominating set problem with the maximum independent set problem. While a maximal independent set is a $5$-ap\-prox\-i\-ma\-tion to both problems, it is easy to obtain a geometric $3$-ap\-prox\-i\-ma\-tion to the maximum independent set problem in $O(n \log n)$ time~\cite{nieberg}. In the graph-based version, a related strategy takes roughly $O(n^5)$ time, though. No similar results are known for the minimum dominating set problem.

The existing PTASs for the minimum dominating set problem in unit disk graphs are based on some packing constraints
that apply to unit disk graphs.\footnote{In \emph{packing problems}, one usually wants to enclose non-{over\-lapping} objects into a recipient  
covering the greatest possible fraction of the recipient area.} 
One of these constraints is the \emph{bounded growth property}: the size of an independent set formed by vertices within distance $r$ of a given vertex, in a unit disk graph, is at most $(1+2r)^2$. Note, however, that the bounded growth property is not tight. For example, for $r=1$, it gives an upper bound of $9$ vertices where the actual maximum size is $5$. Since the bounded growth property is strongly connected to the problem of packing circles in a circle, obtaining exact values for all $r$ seems unlikely~\cite{Fodor}.

\paragraph*{Our contribution}
Our main result consists of the two approximation algorithms given in Section~\ref{s:algorithm}: a graph-based algorithm, which runs in linear $O(n+m)$ time, and its geometric counterpart, which runs in $O(n\log n)$ time in the Real RAM model, regardless of the number of edges.
The approximation factor of both algorithms is $44/9$. The strategy in both cases is to construct a $5$-approximate solution using the algorithm from~\cite{heuristics}, and then perform local improvements to that initial dominating set. Our main lemma (Lemma~\ref{l:irreducible}) uses forbidden subgraphs to show that a solution that admits no local improvement is a $44/9$-approximation. Since the dominating sets produced by our algorithms are independent sets, the same approximation factor holds for the independent dominating set problem.

Proving that a certain graph is \emph{not} a unit disk graph (and is therefore a forbidden induced subgraph) is no easy feat\footnote{The fastest known algorithm to decide whether a given graph is a unit disk graph is doubly exponential~\cite{spinrad}.}. We make use of an assortment of results from discrete geometry in order to prove properties of unit disk graphs that are interesting \textit{per se}.
For example, we use universal covers and disk packings to show that the neighborhood of a clique in a unit disk graph contains at most $12$ independent vertices.
These properties, along with a tighter version of the bounded growth property, are collected in Section~\ref{s:forbidden}, and
allow us to show that certain graphs are not unit disk graphs. Consequently, the analyses of our algorithms employ a broader set of forbidden subgraphs which include, but are not limited to, the $K_{1,6}$.

Additionally, in Section~\ref{s:partial}, we show that a possible, somewhat natural refinement to our graph-based algorithm leads to a tighter $43/9$-approximation, albeit for the price of an extra $O(n^2)$ multiplying factor in the time complexity of the algorithm.

\section{Forbidden subgraphs} \label{s:forbidden}

In this section, we introduce some lemmas about the structure of unit disk graphs. These lemmas will be applied to prove our approximation factors in Sections~\ref{s:algorithm} and \ref{s:partial}. We start by stating three previous results from the area of discrete geometry. The first lemma comes from the study of universal covers  (for a recent survey see~\cite{constants}).
\begin{lem}[P\'al~\cite{pal}]\label{l:universal_cover}
If a set of points $P$ has diameter $1$, then $P$ can be enclosed by a circle of radius $1/\sqrt{3}$.
\end{lem}

Packing congruent disks in a circle is a well-studied problem. Exact bounds on the radius of the smallest circle packing $k$ unitary disks are known for some small values of $k$, namely $k \leq 13$ and $k=19$~\cite{Fodor}. The bound for $k=13$ will be useful to us.
\begin{lem}[Fodor~\cite{Fodor}]\label{l:pack13}
The radius of the smallest circle enclosing $13$ points with mutual distances at least $1$ is $(1+\sqrt{5})/2$.
\end{lem}

The \emph{density} of a packing is the ratio between the covered area and the total area. The following general upper bound is useful when no exact bound is known.
\begin{lem}[Fejes T\'oth~\cite{density}]\label{l:density}
Every packing of two or more congruent disks in a convex region has density at most $\pi/\sqrt{12}$.
\end{lem}

Given a graph $G = (V,E)$ and a vertex $v \in V$, let $N(v)$ denote the \emph{open neighborhood} of $v$ and let $N[v] = N(v) \cup \{v\}$ denote the \emph{closed neighborhood} of~$v$. More generally, the \emph{open $r$-neighborhood} of a vertex $v$ is the set of vertices $w$ such that the distance between $v$ and $w$ in $G$ is exactly $r$, while the \emph{closed $r$-neighborhood} of a vertex $v$ is the set of vertices $w$ such that the distance between $v$ and $w$ in $G$ is at most $r$. For a set $S \subseteq V$, we let $N_{S}(v) = N(v) \cap S$ and $N_{S}[v] = N[v] \cap S$. Finally, given a subgraph $H$ of $G$, the closed neighborhood of $H$, denoted $N[H]$, is the set of vertices that belong to the closed neighborhood of some vertex of $H$. The following two lemmas concern neighborhoods in unit disk graphs.
\begin{lem} \label{l:clique}
The closed neighborhood of a clique in a unit disk graph contains at most $12$ independent vertices.
\end{lem}
\begin{pf}
By Lemma~\ref{l:universal_cover}, the points which define a clique in a unit disk graph can be enclosed by a circle of radius $1/\sqrt{3}$. Therefore, the points corresponding to the closed neighborhood of such a clique are enclosed by a circle of radius $1 + (1/\sqrt{3})$. By Lemma~\ref{l:pack13},
a circle enclosing $13$ points with mutual distances at least $1$ has radius at least $(1+\sqrt{5})/2$. Since $(1+\sqrt{5})/2 > 1 + (1/\sqrt{3})$, the lemma follows.\qed \end{pf}
\begin{lem} \label{l:2neighborhood}
Given an integer $r \geq 1$, the closed $r$-neighborhood of a vertex in a unit disk graph contains at most $\lfloor \pi (2r+1)^2 / \sqrt{12} \rfloor$ independent vertices.
\end{lem}
\begin{pf}
All $n$ disks of diameter $1$ corresponding to the closed $r$-neighborhood of a vertex $v$ must be enclosed by a circle $Z$ of radius $(2r+1)/2$ centered on $v$. Each disk of diameter $1$ has area $\pi/4$ and $Z$ has area $(2r+1)^2 \pi / 4$. Using Lemma~\ref{l:density}, we have $(n\; \pi / 4) / ((2r+1)^2 \pi / 4) \leq \pi/\sqrt{12}$, and the lemma follows.\qed
\end{pf}

We say that a graph $G$ is \emph{$(k,\ell)$-pendant} if there is a vertex $v$ in $G$ with $k$~vertices of degree~$1$ in the open neighborhood of $v$ and $\ell$ vertices of degree~$1$ in the open $2$-neighborhood of $v$. We refer to $v$ as a \emph{generator} of the $(k,\ell)$-pendant graph. The following lemma bounds the value of the parameter $\ell$ for $(4,\ell)$-pendant unit disk graphs.

\begin{lem} \label{l:4l-pendant}
If $G$ is a $(4,\ell)$-pendant unit disk graph, then $\ell \leq 8$.
\end{lem}
\begin{pf}
Let $v$ be a generator of $G$. Since $K_{1,6}$ is a forbidden induced subgraph~\cite{heuristics} and $v$ has $4$ neighbors of degree $1$, we have that the remaining neighbors of $v$ together with $v$ itself form a clique. By Lemma~\ref{l:clique}, it follows that $4 + \ell \leq 12$.\qed
\end{pf}

Next, we consider the case of $(3,\ell)$-pendant unit disk graphs.

\begin{lem} \label{l:3l-pendant}
If $G$ is a $(3,\ell)$-pendant unit disk graph, then $\ell \leq 16$.
\end{lem}
\begin{pf}
Let $v$ be a generator of $G$. Since two vertices are adjacent if and only if their Euclidean distance is at most $1$, the closed neighborhood of $v$ lies inside a circle of radius $1$ centered at $v$. Let $u$ be a neighbor of $v$ with \linebreak degree $1$. We divide the aforementioned circle into six congruent sectors $s_1,\ldots,s_6$ in such a way that $u$ sits on the boundary of two adjacent sectors $s_1,s_2$. Since the diameter of each sector is $1$ and $u$ has degree $1$, we have that $s_1 \cup s_2$ only contains vertex $u$. By the same argument, the remaining two neighbors of $v$ that have degree $1$ are contained in two of the remaining four sectors, which we call $s_3,s_4$ (note that the sectors do not necessarily appear in the order $s_1,\ldots,s_6$). Therefore, only sectors $s_5,s_6$ may contain the neighbors of $v$ that have degree greater than $1$.

Notice that the neighbors of the vertices in $s_5\cup s_6$ should be located within Euclidean distance at most $1$ from $s_5\cup s_6$. Consequently, the disks of unit diameter corresponding to such vertices should be completely contained in the region defined as the Minkowski sum of $s_5 \cup s_6$ and a disk of radius $1{.}5$, i.e.~the region within distance at most $1{.}5$ from $s_5 \cup s_6$. We compute upper bounds to the area of such a region by considering two cases, depending on whether $s_5$ and $s_6$ are opposite sectors (bounded by the same pair of straight lines).

\begin{figure}[t]
 \centering
 \includegraphics[scale = .8]{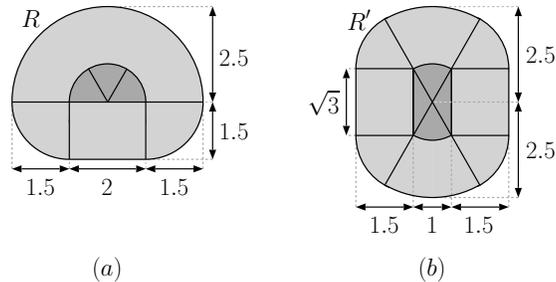}
 \caption{\label{f:sectors}Upper bounds to the area of the Minkowski sum of $s_5 \cup s_6$ and a disk of radius $1{.}5$ in the two different scenarios discussed in the proof of Lemma~\ref{l:3l-pendant}: in (a), the dark area corresponds to a semicircle containing $s_5 \cup s_6$; in (b), it corresponds to the convex hull of~$s_5 \cup s_6$.}
\end{figure}

First, we consider the case when $s_5$ and $s_6$ are \emph{not} opposite sectors. In this scenario, $s_5 \cup s_6$ is contained in a semicircle of radius $1$ centered at the generator $v$, as represented in Figure~\ref{f:sectors}(a). We define a region $R$ as the locus of points within distance at most $1{.}5$ from the aforementioned semicircle of radius $1$. The area $a$ of $R$ is therefore $a = 3 + 17\pi/4$.

By Lemma~\ref{l:density}, the number of disks of unit diameter contained in $R$ is at most $
a \cdot \left(\frac{\pi}{\sqrt{12}} \right) \cdot \left(\frac{1}{\pi/4}\right) < 18{.}8814 < 19.$
Since $k=3$ of these disks are the degree-1 neighbors of $v$, we have $\ell<16$.

Second, we consider the case when $s_5$ and $s_6$ are opposite sectors. In this scenario, the region $s_5 \cup s_6$ is not convex. In order to obtain a convex region $R'$, we define $R'$ as the locus of points within distance at most $1{.}5$ from the convex hull of $s_5 \cup s_6$ (see Figure~\ref{f:sectors}(b)). The area $a'$ of $R'$ is therefore $7\sqrt{3}/2 + 43\pi/12$.

By Lemma~\ref{l:density}, the number of disks of unit diameter contained in $R'$ is at most $a' \cdot \left(\frac{\pi}{\sqrt{12}} \right) \cdot \left(\frac{1}{\pi/4}\right) < 19{.}9989 < 20.$
Since $k=3$ of these disks are the degree-1 neighbors of $v$, we have $\ell<17$.
\qed
\end{pf}

Finally, the following lemma holds for the general case.

\begin{lem} \label{l:kl-pendant}
If $G$ is a $(k,\ell)$-pendant unit disk graph, then $k + \ell \leq 22$.
\end{lem}
\begin{pf}
Immediately from Lemma~\ref{l:2neighborhood} with $r=2$.\qed
\end{pf}

\section{Linear-time 44/9-approximation} \label{s:algorithm}

In this section, we present two 44/9-approximation algorithms. The key property to analyze the approximation factor is presented in Lemma~\ref{l:irreducible}, while the running time analyses are presented in Sections~\ref{s:graph-alg} and~\ref{s:geo-alg}.

Hereafter, let $G = (V,E)$ be a unit disk graph, and let $D \subseteq V$ be an independent dominating set of $G$. If $v \in D$ and $uv \in E$, we say that $v$ \emph{dominates} $u$ and, conversely, that $u$ \emph{is dominated} by $v$.

As already mentioned, unit disk graphs are free of induced $K_{1,6}$. Therefore, at most $5$ vertices of $D$ may belong to the closed neighborhood of any given vertex $v \in V$. A \emph{corona} is a set $C \subseteq D$ consisting of exactly $5$ neighbors of some vertex $c \in V \setminus D$. Such a vertex $c$, which is not necessarily unique, is called a \emph{core} of the corona $C$, whereas the $5$ vertices of the corona are referred to as the corona's \emph{petals}. Notice that the subgraph induced by a corona $C$ and a corresponding core $c$ is a $K_{1,5}$.

Given a dominating set $D$, a corona $C \subseteq D$ is said to be \emph{reducible} if there is a core $c$ of $C$ such that $D \cup \{c\} \setminus C$ is a dominating set. 
We refer to the operation that converts $D$ into the smaller dominating set $D \cup \{c\} \setminus C$ as a \emph{reduction} of $C$ with respect to $c$. If there is no core allowing for a reduction of $C$, than $C$ is dubbed \emph{irreducible}. 
If $C$ is an irreducible corona, then, for every core $c$ of $C$, 
there must exist a vertex $w \in V \setminus (C \cup \{c\})$, such that:
\begin{enumerate}[(i)]
\item $w$ is not adjacent to $c$;
\item $w$ is only dominated, in $D$, by vertices that belong to $C$.
\end{enumerate}
We call $w$ a \emph{witness} of $c$, conveying the idea that the corona having $c$ as a core cannot be reduced with respect to $c$ due to the existence of $w$. 

\begin{lem} \label{l:irreducible}
Let $G = (V,E)$ be a unit disk graph, $D$ an independent dominating set in $G$, and $D^*$ a minimum dominating set of $G$. If all coronas in $D$ are irreducible, then $\rho := |D|/|D^*| \leq 44/9$.
\end{lem}
\begin{pf}We use a charging argument to bound the ratio between the cardinalities of $D$ and $D^*$. Consider that each vertex $u \in D$ splits a \emph{unit charge} evenly among the vertices in the closed neighborhood $N_{D^*}[u]$. The function ${f : D^* \to (0,5]}$ below corresponds to the total charge assigned to each vertex $v^* \in D^*$, accumulating the (fractional) charges that $v^*$ receives from the vertices in $N_D[v^*]$:

\begin{eqnarray}\label{eq:fv}
f(v^*) = \sum_{u \in N_D[v^*]} \frac{1}{\left|N_{D^*}[u]\right|}.
\end{eqnarray}
Note that, since $D$ and $D^*$ are dominating sets, neither $N_{D^*}[u]$ nor $N_{D}[v^*]$ are ever empty, and $f(v^*) \leq |N_D[v^*]|$. Such function $f$ allows us to write the cardinality of $D$ as
$$|D| = \sum_{v^* \in D^*} f(v^*).$$

Since
$$\rho = \frac{|D|}{|D^*|} = \frac{\sum_{v^* \in D^*} f(v^*)}{|D^*|}$$
is precisely the average value of $f(\cdot)$ over the elements of $D^*$, we obtain the desired bound $\rho \leq 44/9$ by showing that the existence of vertices $c^*$ \linebreak in $D^*$ with $f(c^*) > 44/9$ is counterbalanced by a sufficiently large number of vertices $r^*$ in $D^*$ with $f(r^*) \leq 4$.

Before we continue, note that $f(c^*) > 44/9$ means that $f(c^*) = 5$, because the summation in (\ref{eq:fv}) has at most $5$ terms, all of which are of the form $1/i$ for some integer $i \geq 1$.
Thus, let $c^*$ be a vertex in $D^*$ with $f(c^*) = 5$. Clearly, $c^* \notin D$, otherwise $f(c^*) \leq |N_D[c^*]| = 1$, because $D$ is an independent set. Moreover, $c^*$ must have exactly $5$ neighbors in $D$, since a larger number of neighbors in $D$ would imply the existence of an induced $K_{1,6}$ in $G$, which is not possible, and a smaller number would imply $f(c^*) \leq |N_D[c^*]| \leq 4$, a contradiction. Therefore, vertex $c^*$ is a core.

Now let $C \subseteq D$ be the corona of which $c^*$ is a core. Because there are no reducible coronas in $D$, the core $c^*$ must have a witness $w$.
Note that, for all petals $u \in C$, the only vertex in $N_{D^*}[u]$ is the core~$c^*$. Otherwise, the contribution of some $u \in C$ in the summation yielding $f(c^*)$ --- given by (\ref{eq:fv}) --- would be at most $1/2$, and $f(c^*)$ would be at most $9/2 < 5$, a contradiction. In particular, the above implies that the witness $w$, which is adjacent to at least one vertex in $C$, cannot belong to $D^*$. But $D^*$ is a dominating set, so there must exist a vertex $r^* \in D^*$ that is adjacent \linebreak to $w$, and $r^* \neq c^*$ because a witness $w$ is not adjacent to the corresponding core by definition. We call $r^*$ a \emph{reliever} of~$c^*$. Figure~\ref{f:witness} illustrates this situation.

\begin{figure}
 \centering
 \includegraphics[scale = .8]{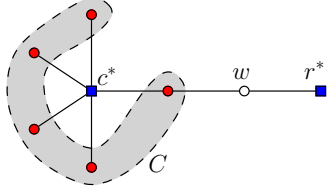}
 \caption{\label{f:witness} Figure for the proof of Lemma~\ref{l:irreducible}. A square indicates a vertex of $D^*$, a solid circle a vertex of the corona $C \subseteq D$, and a hollow circle a vertex not in $D \cup D^*$. Vertices $w$ and $r^*$ are respectively witness and reliever of core $c^*$.}
\end{figure}

We now show that $|N_D[r^*]| \leq 4$. For sake of contradiction, assume \linebreak $|N_D[r^*]| > 4$. Because $G$ is free of induced $K_{1,6}$, such number must be \linebreak exactly $5$, so that $r^*$ is the core of a corona $C' \subset D$. Such a corona must be disjoint from corona $C$, otherwise there would be a vertex in \mbox{$C \cap C'$} adjacent to more than one vertex in $D^*$, namely $c^*$ and $r^*$, contradicting the fact that the only neighbor in $D^*$ of any petal of $C$ is the core~$c^*$. Since, by definition, the witness $w$ is only dominated in $D$ by vertices of $C$, we have $N_{C'}(w) = \emptyset$. Hence, $C' \cup \{w\}$ is an independent set of $G$, constituting, along with the core $r^*$, an induced $K_{1,6}$ in $G$. This is a contradiction, because $G$ is a unit disk graph. Thus, $|N_D[r^*]| \leq 4$. Since $f(r^*) \leq |N_D[r^*]|$, we have $f(r^*) \leq 4$.

We have just shown that the existence of a vertex $c^*$ in $D^*$ with \mbox{$f(c^*) = 5$} implies the existence of a vertex $r^* \in D^*$ such that $f(r^*) \leq 4$. Were this correspondence one-to-one, we would be able to state that the average \linebreak of $f(\cdot)$ over the elements of $D^*$ was no greater than $4{.}5$. Unfortunately, this correspondence is not necessarily one-to-one, as exemplified by the graph in Figure~\ref{f:badgraph}, for which a disk model is given in Figure~\ref{f:badgraph_model} with coordinates presented in Table~\ref{t:coordinates}.

\begin{figure}[t!]
 \centering
 \includegraphics[scale = .8]{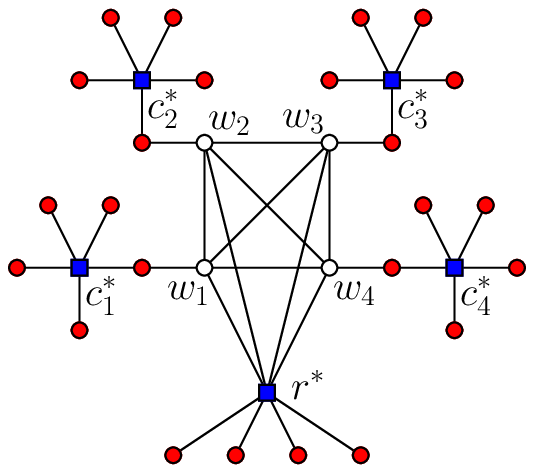}
 \caption{\label{f:badgraph} Unit disk graph where $4$ distinct cores $c^*_1,\ldots,c^*_4$ share the same reliever $r^*$.}
\bigskip
\bigskip
\bigskip
 \centering
 \includegraphics[scale = .26]{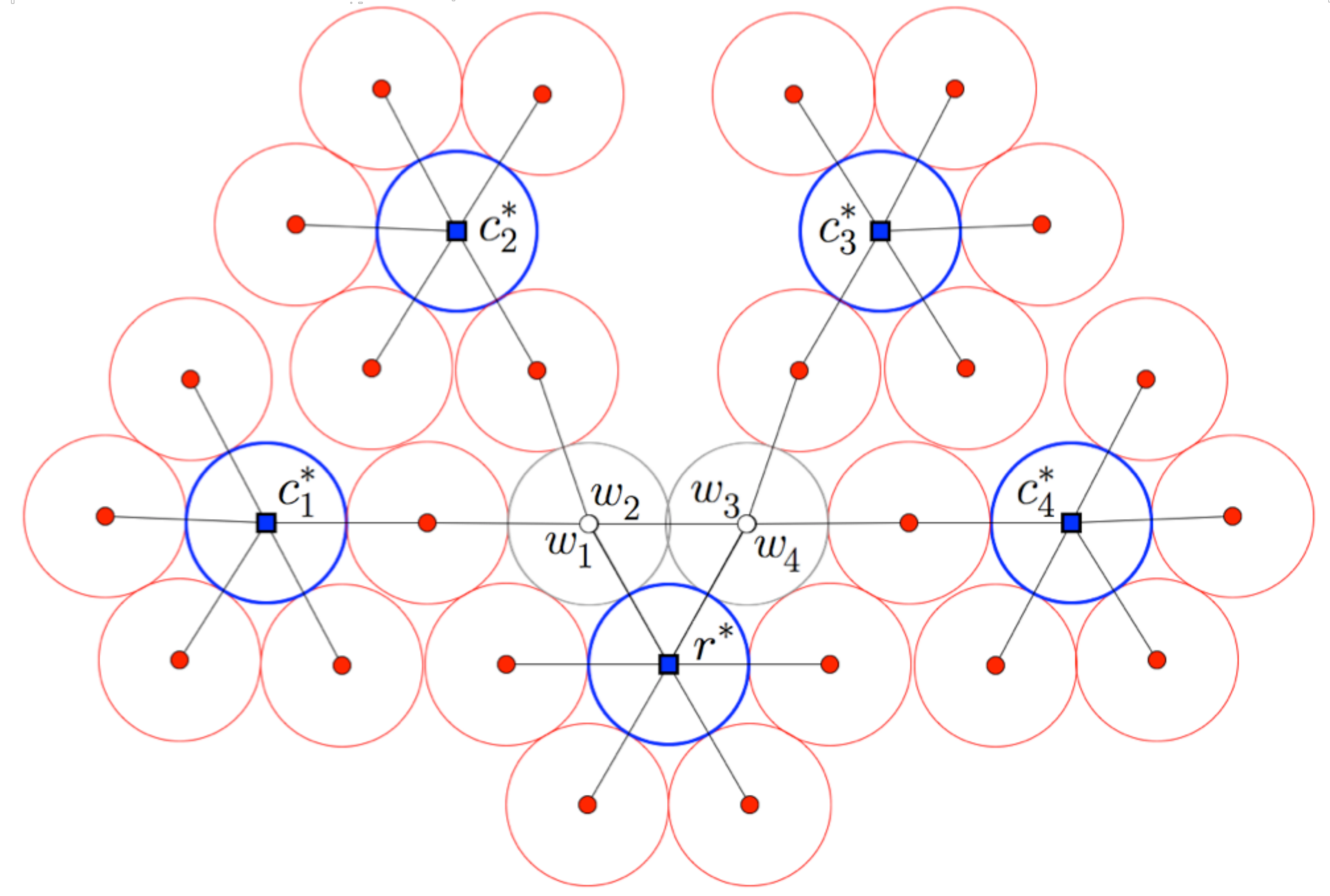}
 \caption{\label{f:badgraph_model} Geometric model for the graph in Figure~\ref{f:badgraph}. Due to scale/precision issues, some disks appear to touch one another when in fact they do not. For clarity, disks that actually touch one another have their centers connected by a straight line. Moreover, there are no concentric disks among those for the witnesses $w_1, \ldots, w_4$. The coordinates of the centers are given in Table~\ref{t:coordinates}.}  
\end{figure}
 
Still, the lemmas in Section~\ref{s:forbidden} allow us to bound the ratio between the number of vertices $c^*$ with $f(c^*) = 5$ and the number of vertices $r^*$ for which the values of $f$ are significantly lower. Let $r^* \in D^*\setminus D$ be a reliever. In order to obtain the claimed bound, we consider two cases depending on the size \linebreak of $N_D[r^*]$:

\smallskip

(i) $\boldsymbol{|N_D[r^*]| \leq 3.}$ By Lemma~\ref{l:2neighborhood}, the closed $4$-neighborhood of $r^*$ contains at most $73$ independent vertices. Since each corona contains $5$ independent vertices, at most $\lfloor (73 - |N_D[r^*]|)/5 \rfloor = 14$ cores may share a common reliever\footnote{We would like to thank an anonymous referee for this simplified argument.}. To derive an upper bound, let $c^*_1,\ldots,c^*_{14} \in D^*$ denote such cores. If $|N_D[r^*]| \leq 3$, then the average value of $f(\cdot)$ among $r^*,c^*_1,\ldots,c^*_{14}$ is at most
 $$\frac{1 \cdot 3 + 14 \cdot 5}{15} < 4{.}867.$$

(ii) $\boldsymbol{|N_D[r^*]| = 4.}$ By Lemma~\ref{l:4l-pendant}, if $|N_D[r^*]| = 4$, then at most $8$ cores $c^*_1, \ldots,c^*_8$ may have $r^*$ as their common reliever, for otherwise we obtain \linebreak a $(4,9)$-pendant graph, which cannot be a unit disk graph. Thus, the average value of $f(\cdot)$ among $r^*,c^*_1,\ldots,c^*_8$ is at most
$$\frac{1 \cdot 4 + 8 \cdot 5}{9} = 44/9 = 4.888\ldots$$

The worst case is $|N_D[r^*]| = 4$, and therefore $\rho \leq 44/9$, concluding the proof.\qed
\end{pf}

\begin{table}
\begin{center}
\begin{tabular}{|lll|}
\hline
 $r^*:(0, 0);$ & & \\
 $c^*_1:(-2492384, 879081),$ & $w_1:(-492423, 870355),$ & \\
 $c^*_2:(-1310377, 2686162),$ & $w_2:(-484809, 874619),$ & \\
 $c^*_3:(1310377, 2686162),$ & $w_3:(484809, 874619),$ & \\
 $c^*_4:(2492384, 879081),$ & $w_4:(492423, 870355);$ & \\
\multicolumn{3}{|l|}{remaining vertices:}\\
~$(\pm776025, 3531423),$ & $(\pm1492384, 879081),$ & $(\pm999986, 5235),$ \\
~$(\pm2309705, 2722805),$ & $(\pm3491646, 917468),$ & $(\pm3023782, 31960),$ \\
~$(\pm1776763, 3570742),$ & $(\pm1840296, 1838114),$ & $(\pm2022913, -3866),$ \\
~$(\pm503019, -864274),$ & $(\pm2957226, 1764474),$ & $(\pm810377, 1820137).$ \\
\hline
\end{tabular}
\caption{\label{t:coordinates} Coordinates of the centers of the disks in Figure~\ref{f:badgraph_model}. All diameters are equal to~$1000001$.}
\end{center}
\end{table}

\subsection{Graph-based algorithm} \label{s:graph-alg}

By Lemma~\ref{l:irreducible}, an independent dominating set with no reducible coronas is a $44/9$-approximation to the minimum dominating set. We now describe how to obtain such a set in linear time given the adjacency list representation of the graph.

We can easily compute a maximal independent set $D$, which is a $5$-approx\-i\-ma\-tion to the minimum dominating set~\cite{heuristics}, in $O(n+m)$ time. An independent dominating set with no reducible coronas can then be obtained by iteratively performing reductions. However, naively performing such reductions leads to a running time of $O(n^2m)$, since (i) there are $O(n)$ candidate cores for a reducible corona, (ii) detecting whether a vertex $v$ is in fact the core of a reducible corona by inspecting the $3$-neighborhood of $v$ takes $O(m)$ time, and (iii) we may need to reduce a total of $O(n)$ coronas. Fortunately, the following algorithm modifies the set $D$ and returns an independent dominating set with no reducible coronas in $O(n+m)$ time.

\begin{enumerate}[(1)]
\item For each vertex $v \in V \setminus D$, compute $N_D(v)$.

\item For each vertex $v \in V \setminus D$, if $|N_D(v)| = 5$, add $N_D(v)$ to the list of coronas $\CC$ (unless it is already there).

\item Let $B \gets \emptyset$. For each corona $C \in \CC$, if there is a vertex $c$ such that $D \cup \{c\} \setminus C$ is a dominating set, then add $c$ to the set $B$.

\item Choose a maximal subset $B'$ of $B$ such that the pairwise distance of the vertices in $B'$ is at least $5$.

\item For each vertex $c \in B'$, perform a reduction $D \gets D \cup \{c\} \setminus N_D(c)$.

\item Repeat all the steps above until $B' = \emptyset$.
\end{enumerate}

The algorithm is correct since all changes made to $D$ along its execution preserve the property that $D$ is an independent dominating set. Notice that, in step (4), we only reduce coronas that are sufficiently far from each other, in order to guarantee that we do not reduce a corona that may have ceased to be reducible due to a previous reduction. Moreover, the algorithm always terminates because the size of $D$ decreases at every iteration, except for the last one. 

Next, we show that the running time is $O(n+m)$. 
Step (1) can be easily implemented to run in $O(n+m)$ time. To execute step (2) in $O(n+m)$ time, we must determine in constant time whether a corona is already in the list $\CC$. This can be achieved by indexing each corona $C$ by an arbitrary vertex $v \in C$ (say, the one with the lowest index), and by storing with $v$ a list of coronas that are in $\mathcal{C}$ and whose index is $v$. Note that, because of the packing constraints inherent to unit disk graphs, the number of coronas that contain a given vertex is constant.

Step (3) can be implemented as follows (for each corona $C \in \CC$):

\begin{enumerate}[(3a)]
\item Let $S_1$ be the union of the open neighborhoods of the $5$ petals of $C$.

\item Let $S_2$ be the set of vertices dominated only be vertices of $C$, i.e., $S$ contains every vertex $w \in S_1$ such that $N_D(w) \subseteq C$.

\item Let $S_3$ be the intersection of the closed neighborhoods $N[v]$ of all $v \in S_2 \cup C$.

\item If $S_3 \neq \emptyset$, then add an arbitrary vertex of $S_3$ to the set $B$.
\end{enumerate}

The steps above take $O(n+m)$ total time when executed for all coronas $C \in \CC$, because the number of coronas that contain or are adjacent to a given vertex is also constant by packing constraints.

It is easy to perform steps (4) and (5) in linear time. It remains to show that the whole process is only repeated for a constant number of iterations. Let $B_1, \ldots, B_k$ denote the set of reducible coronas at each iteration of the algorithm with $B_k = \emptyset$. Note that the reductions performed in step (5) never create a new reducible corona. Therefore $B_1 \supsetneq \cdots \supsetneq B_k$. Let $C$ denote a corona that was reduced in the last iteration $k$. If $C$ was not reduced during a previous iteration $i < k$, then another corona within distance $5$ from $C$ was reduced at that very iteration $i$. Since, again by packing constraints, the maximum number of coronas within constant distance from $C$ is itself a constant, we have $k = O(1)$.

The following theorem summarizes the result from Section~\ref{s:graph-alg}.

\begin{thm} \label{thm:graph-alg}
Given the adjacency list representation of a unit disk graph with $n$ vertices and $m$ edges, it is possible to find a $44/9$-approximation to the minimum dominating set problem in $O(n + m)$ time.
\end{thm}

\subsection{Geometric algorithm} \label{s:geo-alg}

In this section, we describe how to obtain an independent dominating set with no reducible corona in $O(n \log n)$ time given the geometric representation of the graph. The input is therefore a set $P$ of $n$ points. Without loss of generality, we assume that the corresponding unit disk graph is connected (otherwise, we can compute the connected components in $O(n \log n)$ time using a Delaunay triangulation~\cite{cg}). We use terms related to vertices of the graph and to the corresponding points interchangeably. For example, we say a set of points is independent if all pairwise distances are greater than $1$.

We want the points of $P$ to be structured in a suitable fashion. Thus, as a preliminary step, we sort the points by $x$-coordinates and by $y$-coordinates separately (such orderings will also be useful later on), and we partition the points of $P$ according to an infinite grid with unitary square cells by performing two sweeps on the sorted points. Without loss of generality, we assume that no point lies on the boundary of a grid cell. Given $p \in P$, let $\sigma(p)$ denote the grid cell that contains~$p$. We refer to the set of at most $8$ non-empty grid cells surrounding a cell $Q$ as the \emph{open vicinity} of $Q$, denoted $N(Q)$, and to the union of $Q$ and its open vicinity as the \emph{closed vicinity} of $Q$, denoted $N[Q]$. Note that a point $p$ can only be adjacent to points in the closed vicinity of $\sigma(p)$, that is, $N[p] \subset N[\sigma(p)]$. Each point $p \in P$ stores a pointer to its containing cell $\sigma(p)$. Also, each cell stores the list of points it contains and pointers to the cells in its open vicinity. 
Since the graph is connected, the diameter of the point set is at most $n-1$, and thus this whole step can be done in $O(n \log n)$ time.

We are now able to show how to compute a maximal independent set $D$ efficiently. We begin by making a copy $P'$ of $P$, and by letting $D \gets \emptyset$. Then we repeat the two following steps while set $P'$ is non-empty. (i)~Choose an arbitrary point $p \in P'$ and add it to set $D$. (ii)~For each point $p'$ in the closed vicinity of $\sigma(p)$, remove $p'$ from $P'$ if $\|pp'\| \leq 1$. When $P'$ becomes empty, $D$ is an independent dominating set.
This process takes $O(n)$ time due to the two following facts. First, a cell belongs to the closed vicinity of a constant number of cells. Second, the maximum number of points inside a cell with pairwise distances greater than $1$ is also a constant.

We now have that $D$ is a maximal independent set, and therefore \linebreak a $5$-approximation to the minimum dominating set. Next, we show how to modify $D$ in order to produce an independent dominating set with no reducible corona, therefore a $44/9$-approximation to the minimum dominating set. The algorithm mirrors the one in Section~\ref{s:graph-alg}, but each step takes no more than  $O(n \log n)$ time using the geometric representation of the graph.

Since $D$ is an independent set and a grid cell $Q$ has side $1$, a simple packing argument shows that $|D \cap Q| \leq 4$. We store the set $D \cap Q$ in the corresponding cell $Q$. In order to compute $N_D(p)$, it suffices to inspect the at most $36$ points in $D \cap Q$ for $Q \in N[\sigma(p)]$. We can then build a list of coronas in $O(n)$ time (steps (1) and (2) of Section~\ref{s:graph-alg}).

To perform step (3), we need to find out whether there is a core $c$ such that $D \cup \{c\} \setminus C$ is a dominating set, for each corona $C = \{p_1,\ldots,p_5\}$. First, we make $S_1$ the union of $N_D(p_i)$ for $1 \leq i \leq 5$. Then, we let $S_2$ be the subset of $S_1$ containing only the points $p$ with $N_D(p) \subseteq C$. These first two steps are similar to steps (3a) and (3b) in Section~\ref{s:graph-alg}. The remaining sub-steps of step (3) are significantly different, though.

We proceed by making $S_3 = S_2 \cup C$. We need to determine whether there is a point $p \in S_3$ that is adjacent to all points in $S_3$. For each $p \in S_3$, let $\beta(p)$ denote the disk of radius $1$ centered at $p$. Let $R$ denote the convex region defined by the intersection of $\beta(p)$ for all $p \in S_3$. A point $p$ is adjacent to all points in $S_3$ if and only if $p \in R$. We can compute the region $R$ in $O(|S_3| \log |S_3|)$ time using divide-and-conquer in a manner analogous to half-plane intersection~\cite{cg}. We can then test whether each point $p \in S_3$ belongs to the region $R$ in logarithmic time using binary search (remember the points were previously sorted). If there is at least one point $p \in S_3 \cap R$, then we add $p$ to the set $B$. Therefore, the whole step (3) takes $O(n \log n)$ time.

In step (4) of the geometric algorithm, we choose an alternative set $B' \subset B$ which can be computed in $O(n)$ time as follows. For each $p \in B$, we add $p$ to $B'$ and then remove from $B$ all points that are contained in the cells within Euclidean distance at most $4$ of $\sigma(p)$. Since by packing constraints there are $O(1)$ points in the intersection of $D$ and the closed vicinity of a cell, we can easily perform step (5) in $O(n)$ time. Finally, the number of repetitions triggered by step (6) is constant by an argument identical to the one given for the graph-based algorithm.

The following theorem summarizes the result from Section~\ref{s:geo-alg}.

\begin{thm} \label{thm:geo-alg}
Given a set of $n$ points representing a unit disk graph, it is possible to find a $44/9$-approximation to the minimum dominating set problem in $O(n \log n)$ time in the Real RAM model of computation.
\end{thm}

\section{Achieving a 43/9-approximation} \label{s:partial}

In the previous section, a 44/9-approximation was obtained by reducing coronas of a maximal independent set $D$ of graph $G$, that is, by subsequently replacing $5$ petals with $1$ core in $D$ as long as that operation preserved dominance. A natural step to tighten the approximation factor is to allow for \emph{weak reductions}, whereby the $5$ petals of a corona $C$ are removed from the independent dominating set $D$, yet not only is a core $c$ of $C$ inserted into $D$ but also some mutually non-adjacent witnesses of $c$, as long as their number is no greater than $3$ and they dominate all witnesses of $c$. By doing so, the weak reduction of $C$ (with respect to $c$) preserves dominance and still shaves off at least one unit from the size of $D$. If such operation is possible on a corona $C$, then $C$ is said to be \emph{weakly reducible}.
A core $c$ which has $4$ (or more) mutually non-adjacent witnesses is said to be an \emph{overwhelmed} core.\footnote{If $c$ is an overwhelmed core of a corona $C$, then it might be the case that a weak reduction on $C$ with respect to $c$ is still possible. If the subgraph $G[W]$ induced by the set $W$ of witnesses of $c$ admits an independent dominating set $W' \subseteq W$ of size no greater than $3$, then $(D \setminus C) \cup \{c\} \cup W' $ is still an independent dominating set of $G$, and its cardinality is strictly less than $|D|$. However, one cannot decide in (close to) linear time whether such a dominating set exists, and that will not be required by our algorithm.}

We consider the graph-based algorithm presented in Section~\ref{s:graph-alg} with some modifications to cope with weak reductions. The whole modified algorithm becomes:

\begin{enumerate}[(1)]
\item For each vertex $v \in V \setminus D$, compute $N_D(v)$.

\item For each vertex $v \in V \setminus D$, if $|N_D(v)| = 5$, add $C = N_D(v)$ to the list of coronas $\CC$ (unless it is already there), and add $v$ to the list $L_C$ containing the cores of $C$.

\item Let $B$ be an initially empty mapping of cores onto sets of witnesses. For each corona $C \in \CC$, if there is a vertex $c \in L_C$ and an independent set $W_c$ with at most $3$ witnesses of $c$ such that $(D \setminus C) \cup \{c\} \cup W_c$ is a dominating set, then add $(c, W_c)$ to $B$. 

\item Choose a maximal subset $B'$ of the cores in $B$ such that the pairwise distance of the vertices in $B'$ is at least $5$.

\item For each vertex $c \in B'$, perform a (weak) reduction $D \gets (D \setminus N_D(c)) \cup \{c\} \cup W_c$.

\item Repeat all the steps above until $B' = \emptyset$.
\end{enumerate}

The new step (3) can be implemented as follows (for each corona $C \in \CC$):

\begin{enumerate}[(3a)]
\item Let $S_1$ be the union of the open neighborhoods of the $5$ petals of $C$.

\item Let $S_2$ be the set of all vertices $w \in S_1$ with $N_D(w) \subseteq C$.

\item For each core $c \in L_C$, greedily obtain a maximal independent set $W_c$ of $S_2 \setminus N[c]$. If $|W_c| \leq 3$, add $(c, W_c)$ to $B$ and break.
\end{enumerate}

Because each core is evaluated separately as to whether it allows for a weak reduction (whereas in the algorithm of Section~\ref{s:graph-alg} a constant number of set operations per corona was executed), step (3c) dominates the complexity of the whole sequence of steps (1) to (5), with a total $O(nm)$ time for running on all coronas $C \in \CC$. Another important difference, as far as time complexity goes, is that in this modified algorithm the number of iterations of the main loop --- steps (1) to (5) --- is no longer $O(1)$, due to the fact that new (weakly) reducible coronas can be created, as illustrated in Figure~\ref{f:newcoronas}. However, the number of iterations is certainly $O(n)$, because the size \linebreak of $D$ decreases by at least $1$ in each iteration. Hence the overall time complexity of the modified algorithm is $O(n^2 m)$.

Next, we establish the approximation factor of the modified algorithm. Note that step (3c) asserts that the coronas that are not (weakly) reduced by the algorithm present only overwhelmed cores.

\begin{lem} \label{l:irreducible_partial}
Let $G = (V,E)$ be a unit disk graph, $D$ an independent dominating set in $G$, and $D^*$ a minimum dominating set of $G$. If all coronas \linebreak in $D$ have only overwhelmed cores, then $\rho = |D|/|D^*| \leq 43/9$.
\end{lem}

\begin{pf}
We follow a strategy similar to that in the proof of Lemma~\ref{l:irreducible} (also using the concept of reliever defined therein): by employing the same function
 ${f : D^* \to (0,5]}$ defined in (\ref{eq:fv}),
we prove there is an appropriate balance among vertices
with high ($> 43/9$) and low ($\leq 4$) images under $f$, thus yielding an average value for $f(\cdot)$ that is no greater than $43/9$ --- the claimed approximation factor.

\begin{figure}
 \centering
 \includegraphics[scale = .82]{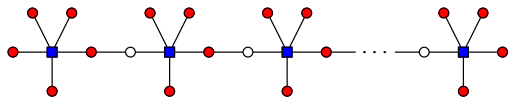}
 \caption{\label{f:newcoronas}Example of graph with a maximal independent set (represented by solid circles) containing only $1$ weakly reducible corona. After every weak reduction performed by the modified algorithm, except for the last one, a new weakly reducible corona is created.}
\end{figure}

Let $c^*$ be a vertex in $D^*$ with $f(c^*) > 43/9$. Because $43/9 > 4.5$, and all (at most $5$) terms in the summation that yields $f$ are either $1$ or no greater than $0.5$, we have that $f(c^*) > 43/9$ implies $f(c^*) = 5$. So $c^*$ is a core.
By hypothesis, $c^*$ is overwhelmed. Hence, $c^*$ possesses at least $4$ mutually non-adjacent witnesses $w_1, \ldots, w_4$, each one implying the existence of a reliever $r^* \in D^*$ with $f(r^*) \leq 4$. Surely, such relievers need not be all distinct, and different cores may share a common reliever.
Still, geometric properties of unit disk graphs allow us to derive upper bounds for the core-to-reliever ratio. We can thus obtain an upper bound to the average value of $f(\cdot)$ in $D^*$, and consequently to the approximation factor of the algorithm.

Suppose there are $t$ cores $c^*_i \in D^*$ such that $f(c^*_i) = 5$. For $i = 1, \dots, t$, we let $C_i$ be the corona having $c^*_i$ as a core, $W_i$ be a set of (at least 4) mutually non-adjacent witnesses of $c^*_i$, and $R_i$ be the set of relievers of $c^*_i$. Now, we construct a bipartite multigraph $G'=(C' \cup R', E')$ as follows. The parts of $G'$ are $C'=\{c^*_1,...,c^*_t\}$ and $R'=R_1\cup...\cup R_t$. The multiset $E'$ contains, between each core $c^*_i \in C'$ and reliever $r^*_j \in R'$, a number $\phi(c^*_i, r^*_j)$ of parallel edges that is equal to the number of petals of $C_i$ adjacent to witnesses $w$ \linebreak (of $c^*_i$) such that $w$ is a neighbor of $r^*_j$.
Thus, the total number of edges incident to a core $c^*_i \in C'$ is

$$d(c^*_i) = \sum_{r^*_j \in R'} \phi(c^*_i, r^*_j).$$

Analogously,
the total number of edges incident to a reliever $r^*_j \in R'$ is

$$d(r^*_j) = \sum_{c^*_i \in C'} \phi(c^*_i, r^*_j).$$

We now obtain an upper bound $\rho' = 43/9$ for the average value of $f(\cdot)$ \linebreak over $C' \cup R'$. Observe that $C'$ contains all vertices $c^*$ of $D^*$ such \linebreak that $f(c^*) > 43/9$, hence the average value $\rho$ of $f(\cdot)$ over the whole set $D^* \supseteq C' \cup R'$ cannot be any greater.

Of course the average value we are interested in depends on the core-to-reliever ratio $|C'| / |R'|$ in $G'$: the more cores (respectively, the fewer relievers) in $C'$ (in $R'$), the greater the average. Therefore, in order to obtain the desired upper bound $\rho'$, we must consider the case in which the elements of $C'$ (respectively, of $R'$) have degrees in $G'$ that are as low (as high) as possible.

First, notice that if $c^*$ is a core of corona $C$ and $f(c^*) = 5$, then it is not possible that more than two non-adjacent witnesses of $c^*$ sharing a common reliever $r^*$ are adjacent to the same petal $p \in C$. Otherwise, let $w_1, w_2$ and $w_3$ be such witnesses. Since $p$ and $r^*$ are non-adjacent (due to the image of $c^*$ under $f$ being $5$), the subgraph of $G$ induced by $\{p, r^*,w_1,w_2,w_3\}$ is a $K_{2,3}$. This is a contradiction, because the graph $K_{2,3}$ is not a unit disk graph~\cite{vanLeeuwen}. Thus, since $c^*$ has at least four witnesses, the number of petals of $C$ adjacent to witnesses of $c^*$ --- and therefore the degree of each core in $C'$ --- is at least $\lceil 4/2 \rceil= 2$. 

The above lower bound can be improved for cores $c^* \in C'$ having a \mbox{reliever $r^*$} with exactly four neighbors in $D$. If a reliever $r^*$ has four neighbors in (the independent set) $D$, then the remaining neighbors of $r^*$ form a clique. Since there is a witness $w$ of $c^*$ among these remaining neighbors of $r^*$, then the other (at least) three mutually non-adjacent witnesses of $c^*$ must have relievers distinct from $r^*$. Moreover, by an argument analogous to the one used in the previous paragraph, those witnesses must be adjacent to at least $\lceil 3/2 \rceil= 2$ petals of the corona having $c^*$ as a core. This means there are at least two edges connecting $c^*$ to its relievers in $R' \setminus \{r^*\}$, plus one edge connecting $c^*$ to $r^*$. Hence, the degree \linebreak of $c^*$ in $G'$ is at least $3$.

The maximum degrees of vertices $r^* \in R'$ depend on the number $|N_D(r^*)|$ of neighbors of $r^*$ in $D$, and now we employ some of the geometric lemmas of Section~\ref{s:forbidden} to infer suitable upper bounds. Recall that any set of petals is an independent set.
\begin{itemize}
\item If $|N_D(r^*)| = 4$ (implying $f(r^*) \leq 4$), then, by Lemma~\ref{l:4l-pendant}, there are at most $8$ petals in the $2$-neighborhood of $r^*$. Hence, $d(r^*) \leq 8$.
\item If $|N_D(r^*)| = 3$ (implying $f(r^*) \leq 3$), then, by Lemma~\ref{l:3l-pendant}, there are at most $16$ petals in the $2$-neighborhood of $r^*$. Hence, $d(r^*) \leq 16$.
\item If $|N_D(r^*)| = 2$ (implying $f(r^*) \leq 2$), then, by Lemma~\ref{l:kl-pendant}, there are at most $20$ petals in the $2$-neighborhood of $r^*$. Hence, $d(r^*) \leq 20$.
\item If $|N_D(r^*)| = 1$ (implying $f(r^*) \leq 1$), then, by Lemma~\ref{l:kl-pendant}, there are at most $21$ petals in the $2$-neighborhood of $r^*$. Hence, $d(r^*) \leq 21$.
\end{itemize}

For $k = 1, \ldots, 4$, let $n_k$ denote the number of relievers in $G$ containing exactly $k$ neighbors in $D$, so that $|R'| = \sum_{k=1}^4 n_k$. Let also $C'_4 \subseteq C'$ be the set of cores having at least one reliever with exactly $4$ neighbors in $D$,\linebreak and $C'_3 = C' \setminus C'_4$ be the set of cores whose relievers have at most three neighbors in $D$. Finally, we let $t_3 = |C'_3|$ and $t_4 = |C'_4|$, so that $t = |C'| = t_3 + t_4$.

Since $G'$ is bipartite, the number of edges incident to $C'$ and the number of edges incident to $R'$ are the same. Consequently, 

$$\sum_{c^* \in C'} d(c^*) = \sum_{r^* \in R'} d(r^*),$$ and
\begin{eqnarray}\label{eq:2}
2 t_3 + 3 t_4 \leq 8 n_4 + 16 n_3 + 20 n_2 + 21 n_1.
\end{eqnarray}

The same reasoning holds for the subgraph of $G'$ induced by the cores \linebreak in $C'_3$ and their relievers in $R'$, so
\begin{eqnarray}\label{eq:3}
2 t_3 \leq 16 n_3 + 20 n_2 + 21 n_1.
\end{eqnarray}
By dividing both sides of (\ref{eq:3}) by $2$ and adding it to (\ref{eq:2}), we obtain
\begin{eqnarray}\label{eq:4}
3(t_3 + t_4) \leq  8 n_4 + 24 n_3 + 30 n_2 + \frac{63}{2} n_1.
\end{eqnarray}
As for the average $\rho'$ of $f(\cdot)$ over the elements of $G'$, we can write

$$\rho' = \frac{5 (t_3 + t_4) + 4 n_4 + 3 n_3 + 2 n_2 + n_1}{t_3 + t_4 + n_4 + n_3 + n_2 + n_1}.$$

Now, substituting $(t_3 + t_4)$ in the expression above by the upper bound obtained from (\ref{eq:4}), we have

$$\rho' \leq \frac{\frac{52}{3} n_4 + 43 n_3 + 52 n_2 + \frac{107}{2} n_1}{\frac{11}{3} n_4 + 9 n_3 + 11 n_2 + \frac{23}{2} n_1}.$$

Thus, $\rho'$ is bounded by a multivariate rational function whose maximum is $43/9 = 4{.}777\ldots$, achieved when $n_1=n_2=n_4 = 0$. 
\qed

\end{pf}

The following theorem summarizes the result from Section~\ref{s:partial}.

\begin{thm} \label{thm:graph-alg-partial}
Given the adjacency list representation of a unit disk graph with $n$ vertices and $m$ edges, it is possible to find a $43/9$-approximation to the minimum dominating set problem in $O(n^2 m)$ time.
\end{thm}

\section{Conclusion and open problems} \label{s:conclusion}

We introduced novel efficient algorithms for approximating the minimum dominating set and minimum independent dominating set in unit disk graphs.

On one hand, a linear-time algorithm was devised attaining a sub-5 approximation factor, namely $44/9 < 4{.}889.$ Nevertheless, the best lower bound we know for the proposed algorithm is $4{.}8$, which corresponds to the unit disk graph given in Figure~\ref{f:badgraph}.
Closing this gap would likely require the development of new tools to prove that certain graphs are not unit disk graphs, for which computer generated proofs may be useful.

\begin{figure}[t!]
 \centering
 \includegraphics[scale = .25]{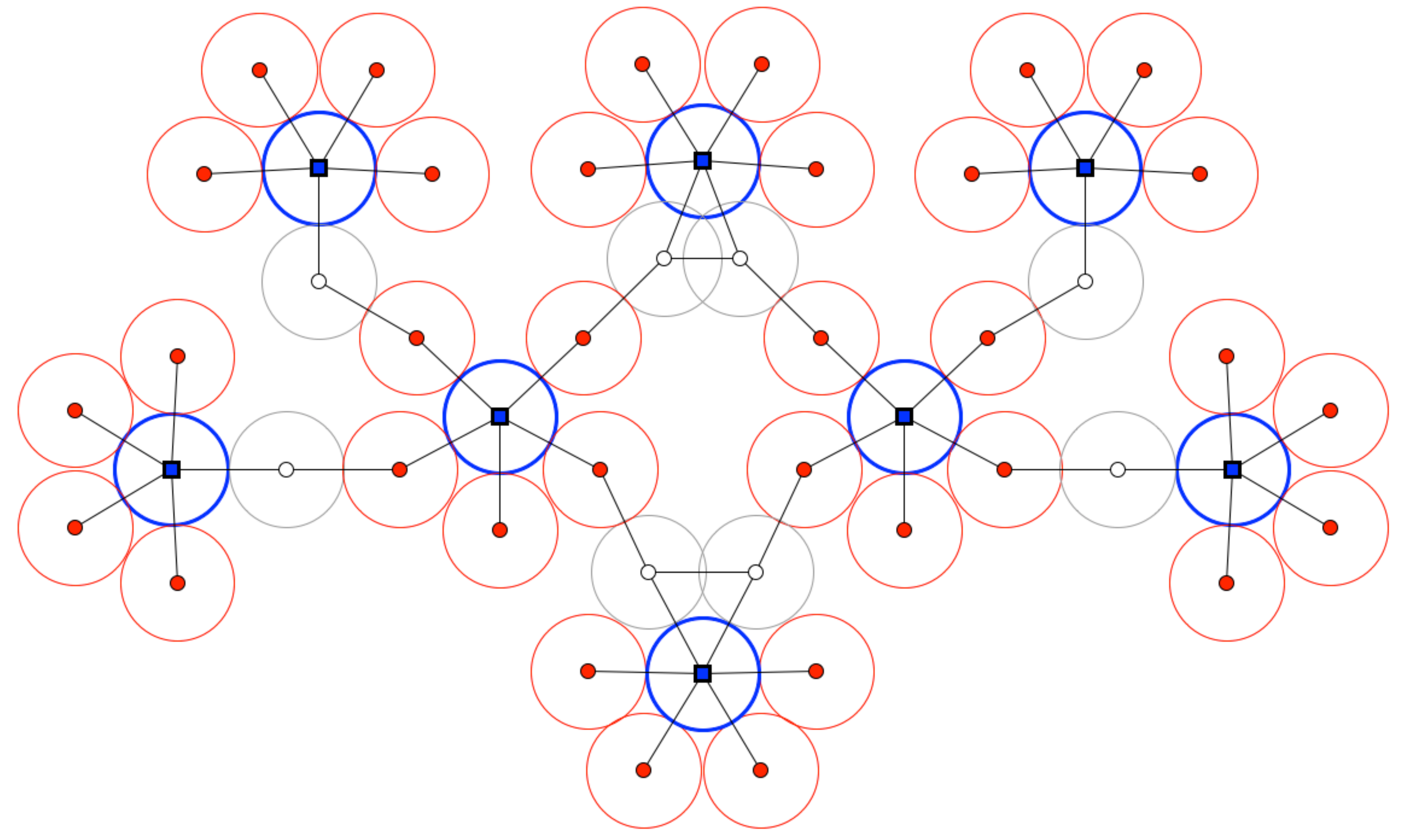}
 \caption{\label{f:lowerbound425}Lower bound of $4{.}25$ to the approximation factor of the modified graph-based algorithm. The coordinates of the centers are given in Table~\ref{t:coordinates2}.}  
\bigskip
~
\end{figure}

\begin{table}[h!]
\begin{center}
\begin{tabular}{|lll|}
\hline
~$(0,0),$ & $(0,4500000),$ & \\
~$(\pm336577, 3647829),$ & $(\pm3372414, 3440722),$ & $(\pm3657983, 1789254),$ \\
~$(\pm469471, 882947),$ & $(\pm2857376, 5297889),$ & $(\pm3887452, 5297889),$ \\
~$(\pm1043683, 2940723),$ & $(\pm2506389, 2940723),$ & $(\pm892089, 1789254),$ \\
~$(\pm2657983, 1789254),$ & $(\pm1775036, 1258725),$ & $(\pm529919, 5348048),$ \\
~$(\pm997564, 4430244),$ & $(\pm4605648, 790625),$ & $(\pm5515150, 1274216),$ \\
~$(\pm5515150, 2304292),$ & $(\pm4605648, 2787883),$ & $(\pm1775036, 2258725),$ \\
~$(\pm2373785, 4388387),$ & $(\pm4657983, 1789254),$ & $(\pm3372414, 4440722),$ \\
~$(\pm515038, -857167),$ & $(\pm999780, 20942),$ & $(\pm4371043, 4388387).$\\
\hline
\end{tabular}
\caption{\label{t:coordinates2} Coordinates of the centers of the disks in Figure~\ref{f:lowerbound425}. All diameters are equal to~$1000001$.}
\end{center}
\end{table}

On the other hand, an enhanced approximation factor of $43/9 < 4{.}778$ was obtained by allowing for more local replacements, yet a lower bound of~$4{.}25$, corresponding to the unit disk graph given in Figure~\ref{f:lowerbound425} (with coordinates in Table~\ref{t:coordinates2}), is the best we are aware of.  
Notwithstanding the $O(n^2)$ factor increase on its time complexity,
such a modified algorithm is still much faster than, say, the state-of-the-art 4-approximation algorithm from~\cite{cccg}, which requires a geometric model as input and runs in $O(n^9)$ time. Moreover, since the number of (weak) reductions that are performed remains linear, it may be possible to conceive either a refined analysis or a smarter implementation.

\bibliographystyle{elsarticle-num}

\end{document}